\def\year{2018}\relax
\pdfoutput=1
\documentclass[letterpaper]{article} 
\usepackage{aaai18} 
\usepackage{times} 
\usepackage{helvet} 
\usepackage{courier} 
\usepackage{url} 
\usepackage{graphicx} 
\usepackage{amsthm}
\usepackage{subcaption}
\usepackage{times}
\graphicspath{ {figures/} }
\begin{document}

\title{When Politicians Talk About Politics: \\ Identifying Political Tweets of Brazilian Congressmen}
\author{Lucas S. Oliveira 
        \\Federal University of Minas Gerais
        \\\texttt{lsoliveira@uesb.edu.br}
        \And Pedro O. S. Vaz de Melo
        \\Federal University of Minas Gerais
        \\\texttt{olmo@dcc.ufmg.br}
        \\\AND {\bf Marcelo S. Amaral}
        \\State University of Southwestern of Bahia
        \\\texttt{amaral69@gmail.com}
        \\\And {\bf Jos{\'e} Ant{\^o}nio. G. Pinho}
        \\Federal University of Bahia 
        \\\texttt{jagp@ufba.br}
        }

\maketitle
\begin{abstract}
Since June 2013, when Brazil faced the largest and most significant mass protests in a generation, a political crisis is in course. In midst of this crisis, Brazilian politicians use social media to communicate with the electorate in order to retain or to grow their political capital. The problem is that many controversial topics are in course and deputies may prefer to avoid such themes in their messages. To characterize this behavior, we propose a method to accurately identify political and non-political tweets independently of the deputy who posted it and of the time it was posted. Moreover, we collected tweets of all congressmen who were active on Twitter and worked in the Brazilian parliament from October 2013 to October 2017. To evaluate our method, we used word clouds and a topic model to identify the main political and non-political latent topics in parliamentarian tweets. Both results indicate that our proposal is able to accurately distinguish political from non-political tweets. Moreover, our analyses revealed a striking fact: more than half of the messages posted by Brazilian deputies are non-political.
\end{abstract}

\section{Introduction}

Democracy is facing a difficult time in Brazil, where a political crisis is in course. It all started in 2013, when Brazil faced the largest and most significant mass protests in a generation \cite{saad2013mass}. One year after the protests, a new general election took place, when Dilma Rousseff was reelected president of Brazil in midst of corruption scandals that involved not only the executive power, but also legislators and companies in a corruption scheme of bribes, kickbacks and inflated contracts \cite{Watts2016}. In the same year, \textit{Opera\c{c}{\~a}o Lava Jato} [Car Wash Operation] started, revealing the biggest corruption scandal in the history of Brazil. Among other things, billionaires were put in jail and a former president was dragged into court. In December 2015 Rousseff was charged of taking loans from state banks without congressional approval \cite{BUCCI2016}, resulting in an impeachment process that was approved by Congress in August 2016.

There are several factors that contribute to the troubled political scenario in Brazil. In short, Brazil is the fifth largest country in the world, has a plethora of political parties~\cite{VazdeMelo2015}, three levels of government and an open-list election system for lawmakers~\cite{mainwaring1999rethinking}. As a consequence, no single party has ever come close to a commanding majority in Congress, so support is bought with cabinet posts and/or cash, and the election process always leads to a patronage or a personalist relation between politicians and their constituents \cite{moises2011desempenho}.

In such systems, social media can play a determinant role. In fact, some studies have shown that social media play an important role in shaping political discourse and can be a valid indicator of offline behavior \cite{Conover2011,amaral2016Tuitando,Lietz2014}. Unfortunately, despite all these efforts, little is known about the content of the speech given by deputies in social media. Defining the correct communication strategy in social media is a major challenge for politicians. Such strategy is in the spectrum of being totally conservative, where the politician shares only non-political, personalistic and individualist communications, and being fully transparent, where the politician fully discloses her/his political views over the current national issues.

Thus, the goal of this work is to accurately classify tweets posted by politicians into two categories: political and non-political. Our method is independent of the deputy who posted the tweet and of the time the tweet was posted. To evaluate our proposal, we collected tweets of all congressmen who were active on Twitter and worked in the Brazilian parliament from October 2013 to October 2017. During this time, Brazil faced two major political events: the 2014 elections and the 2016 impeachment process. The objective of this method is to identify communications focused on meaningful political opinions from those focused on trivialities, such as personalistic and individualistic messages.
 
\section{Related Work}
\label{sec:related}

The task of identifying social media messages about political issues is not new in the literature. Paul \shortcite{Paul} proposed a semi-supervised approach to distinguish tweets related to politics from non-political ones. From a collection of news articles, the authors used a topic model approach to identify words related to political topics. Conover et al.\shortcite{Conover2011} performed a simple hashtag co-occurrence discovery procedure to identify political tweets.
From a collection of tweets posted around the US election day, Gao et al. \shortcite{Gao2017} proposed a weakly supervised two-path bootstrapping approach for detecting and characterizing online hate speech. In the aforementioned studies~\cite{Paul,Conover2011,Gao2017}, a initial list of hashtags or words was used as input for the creation of the final list of terms. Thus, these approaches are very difficult to be generalized or to be accurate over a long period of time. Nevertheless, the neural network architecture proposed by \cite{Gao2017} is similar to the one we propose in this article. However, we used a Convolutional Neural Network (CNN) with different filter size instead of a Long Short Term Memory network (LSTM), as the latter presented worse results in our experiments.

\section{Political Dataset}
\label{sec:political_dataset}
In this work we collected 1.1 millions public tweets from 692 Brazilian deputies from October 2013 to October 2017 by means of the Twitter API (Application Programming Interface). The names of the active congressmen during this period were retrieved and validated by a researcher in March 2015 from the Chamber of Deputies Open Data website\footnote{\url{https://dadosabertos.camara.leg.br}}. The list of the Twitter accounts associated with the congressmen was collected from the personal profile pages of each congressman. After this process, each account was manually validated. Table~\ref{tab:data} summarizes our data set.The tweets and the number of followers and followees of each congressman were collected twice, once in December 2015 and once in November 2017. After that, we merged these two datasets and prepared the text of the tweets for processing: we removed duplicated tweets, punctuation, words with less than 2 characters, Portuguese stop words, hashtags, URLs, and mentions. Finally, we labeled each congressmen according to their position before and after the 2014 elections. If a congressman had a sit in Congress before the elections and was able to be reelected, we labeled they as \textbf{reelected}. Conversely, if a politician had a sit in Congress before the elections and was not able to be reelected, we label they as \textbf{loser}. Finally, if a politician did not have a sit in Congress before the elections and was able to be elected in 2014 (or was a supplant of a congressman after election), we label they as \textbf{newcomer}.
\vspace{-0.3cm}
\begin{table}[h!]
\centering
\caption{Dataset summary}\vspace{-0.2cm}
\label{tab:data}
\resizebox{\linewidth}{!}{
\begin{tabular}{lrrrrr}
\hline
	&	\textbf{\# tweets}	&	\textbf{\# deputies} & \textbf{average} & \textbf{\% political tweets} & \textbf{\# political tweets} 	\\
\hline
Reelected	&	567.565	&	273	& 2.079 & 50 & 283.782 \\
Newcomers	&	320.283	&	202	& 1.586 & 43 & 137.722\\
Losers	&	308.598	&	217	& 1.422 & 43 & 132.697\\
\hline
\textbf{total}	&	1.196.446	&	692 & 1.729 & 139 & 554.201	\\
\hline
\end{tabular}
}
\end{table}
\vspace{-0.3cm}

\section{Identification of Political Tweets}
\label{sec:methodology}

Federal deputies are elected by the population of a country and their duty is to propose, discuss and pass laws, which can change even the Constitution. In this section we present a methodology to identify tweets that have a political content, i.e., tweets that reveal to the population the work or the political view a given deputy has. To the best of our knowledge, we are the first to use this general classification approach to tackle this problem. More formally, we define a political tweet as follows.

\newtheorem{p_def}{Definition}
\begin{p_def}
\label{def1}
A political tweet is a message posted in Twitter by a politician whose content express subjects related to fundamental issues of state, politics, govern and justice. More specifically, such tweets can cover one or more of the following topics: federal programs, projects and laws; political campaign; public statements about the Congress agenda; government subsidies; judicial decisions; public expenditures and crimes against public administration.
\end{p_def}

The process of identifying political tweets using a classification approach involves several methodological decisions. These decisions can be thought as the \textit{meta parameters} of the methodology and are related to the following challenges: (i) the number and the selection of instances to manually label; (ii) the text embedding method to be used to transform tweets into vectors; (iii) the selection of the classification method. Table~\ref{tab:parameters} describes all the meta parameters and their possible values. During our experiments, we verified that the meta parameters are independent among themselves, e.g., changing the text embedding technique does not alter the relative performance of the classification methods. Because of that, exception made to meta parameter being evaluated, the configuration used to generate the results is: 2000 labeled tweets, \textit{Word2Vec C-BoW} with 300 dimensions as the text embedding technique and Convolutional Neural Network (CNN) as the classification method. 
\vspace{-0.1cm}
\begin{table}[h!]
\centering
\caption{Meta Parameters}\vspace{-0.3cm}
\label{tab:parameters}
\resizebox{\linewidth}{!}{
\begin{tabular}{rllrl}
\hline
\textbf{labeled tweets}	&	\textbf{period}	&	\textbf{embedding}	&	\textbf{embedding size}	&	\textbf{classification method}	\\
\hline
100	&	random	&	Word2Vec C-BoW	&	100	&	CNN	\\
500	&	few months	&	Word2Vec Skip-Gram	&	300	&	LSTM	\\
1000	&	few deputies	&	Glove	&		&	FastText	\\
2000	&		& Word2Vec C-BoW over our dataset	&		&		\\
\hline
\end{tabular}
}
\end{table}
\vspace{-0.3cm}

The first challenge is to select the tweets to be manually labeled as \textit{political} and, conversely, \textit{non-political}. Then, we generate classification results for the following number of labeled instances: 100, 500, 1000 and 2000. For all cases, half of the tweets are manually labeled as \textit{political} and half as \textit{non-political}. The collections of tweets to be manually labeled were sampled from the data set using the original frequency distribution of tweets over time. To do that, we grouped all tweets by the month it was posted and got the tweet frequency for each month. Then we calculated how many tweets per month were necessary to sample for each class in order to mimic the original distribution.

\begin{figure*}[h]
\centering
\begin{subfigure}[b]{0.18\textwidth}
 \includegraphics[width=\linewidth]{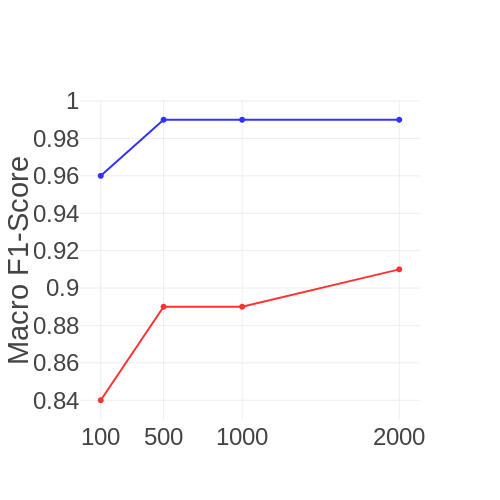}
 \caption{\scriptsize{Sample size}}\label{fig:sample}
\end{subfigure}
\begin{subfigure}[b]{0.18\textwidth}
 \includegraphics[width=\linewidth]{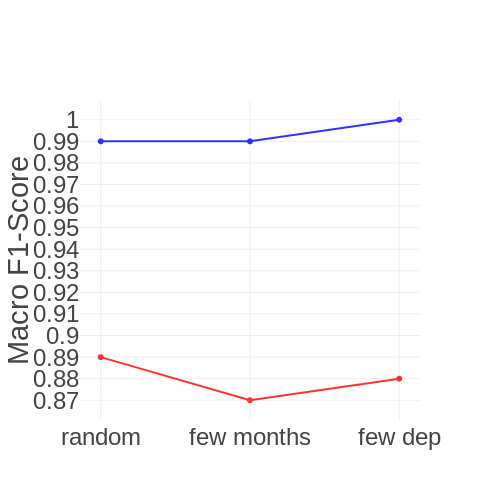}
 \caption{\scriptsize{Dispersion}}\label{fig:dispersion}
\end{subfigure}
\begin{subfigure}[b]{0.18\textwidth}
 \includegraphics[width=\linewidth]{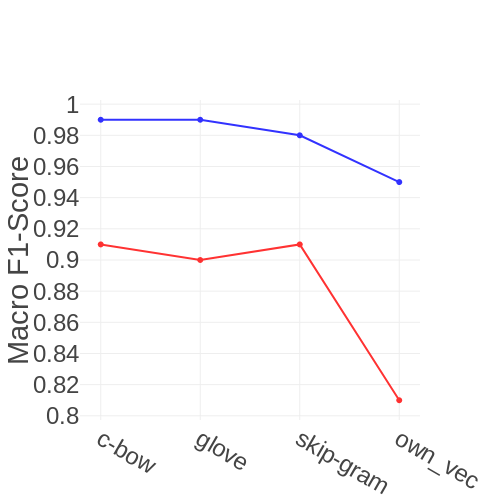}
 \caption{\scriptsize{Embeddings}}\label{fig:embedding}
\end{subfigure}
\begin{subfigure}[b]{0.18\textwidth}
 \includegraphics[width=\linewidth]{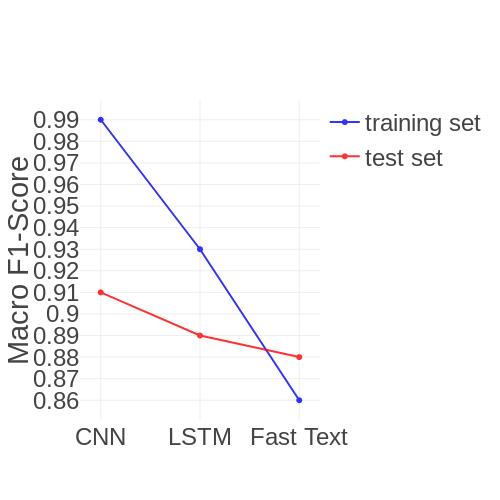}
 \caption{\scriptsize{Classification method}}\label{fig:classification}
\end{subfigure}
\begin{subfigure}[b]{0.18\textwidth}
 \includegraphics[width=\linewidth]{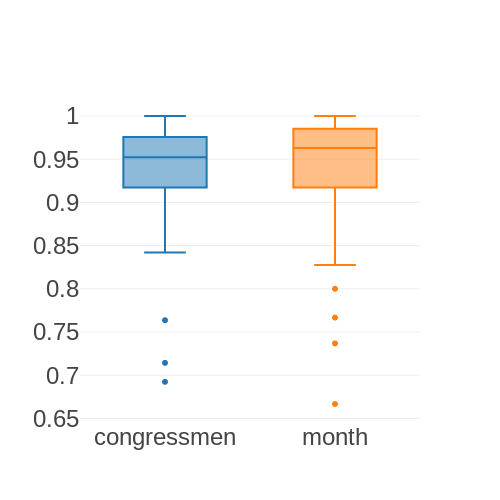}
 \caption{\scriptsize{Bias analysis}}\label{fig:accuracy}
\end{subfigure}
\caption{Classification results. Macro F1 scores for different configurations.}\label{fig:results}\vspace{-0.3cm}
\end{figure*}

Figure \subref{fig:sample} shows the Macro F1 score for the classification task when the size of the manually labeled data is varied. Observe that, as expected, the model accuracy grows as we increase the training set size. Also, observe that, although the cross validation results stabilizes with a training set of 500 instances, the results for the test set grows significantly up to 2000 instances, with a Macro F1 score of 99\% in the training set and 91\% in test set.

It is also worth mentioning the importance of having an unbiased training set in terms of time. To show that, we compared the performance of the classifier when three different training sets are used: (i) the previous randomly and unbiased collection of 500 manually labeled tweets, (ii) a biased collection of 500 labeled tweets in the time dimension and (iii) a biased collection of 500 labeled tweets in the deputy dimension. In these two biased collections, we artificially made the frequency of tweets more skewed towards a few months and, for the second case, a few deputies. In such cases, more than half of the labeled tweets are from only a few months, for the first biased collection, and from a few deputies, for the second biased collection. In Figure \subref{fig:dispersion} we show the Macro F1 score for these three collections of training sets. Observe that for the biased collections, the cross validation results grow, revealing an overfitting. On the other hand, the results in the test set for the unbiased collection decay, what is expected. 

Before running the classification methods, we execute a text embedding technique to transform every word in a numerical vector. We compare four text embedding techniques. The first three word vectors are publicly available and were trained over a large Portuguese data set \cite{Hartmann2017}, which is able to produce an embedding matrix for a vocabulary of 1.3 trillions words. These vectors were produced using the following methods: \textit{Word2Vec C-BoW}~\cite{Mikolov2013}, \textit{Word2Vec Skip-Gram}~\cite{Mikolov2013} and \textit{Glove}~\cite{pennington2014glove}. Additionally, we trained the Word2Vec C-BoW model using the collection of tweets of deputies described in this work.

Thereafter, we evaluated the different embedding techniques using the parameters described in Table~\ref{tab:parameters}. Figure \subref{fig:embedding} exhibits that \textit{Word2Vec C-BoW} and \textit{Glove} have the same 99\% of accuracy in the training set. On the other hand, the result in the test set show that \textit{Word2Vec C-BoW} achieved a higher Macro F1 score than \textit{Glove}. Also, it is important to note that\textit{Word2Vec Skip-Gram} and \textit{Word2Vec C-BoW} obtained the same 91\% of Macro F1 score in test set. Finally, the \textit{Word2Vec} model trained using our dataset obtained the worst results.

The last decision is to choose which Neural Network architecture to use. More specifically, we evaluated three different architectures: Convolutional Neural Network (CNN) \cite{Kim2014}, Long Short Term Memory networks (LSTM) \cite{Hochreiter} and FastText \cite{Joulin2016}. Evaluation was done through a 10-Fold Cross Validation in the training set and, after that, we calculated the Macro F1 scores. In addition, we also validated the result in a external test set using the same Macro F1 score. 

For comparison purposes, we standardize the neural network input layer and output layer. In the input layer each word in a congressman tweet is represented as a dense vector with 300 dimensions learned by \textit{Word2Vec C-BoW}. 
In case the word is not present in the vocabulary, we replaced it by a special symbol UNK (unknown) and get its embedding representation. Thus, we have a matrix of words and embeddings with vocabulary size $\times$ 300 dimensions that we provide as the embedded input layer. Additionally, for the output layer, we use a single neuron with a sigmoid activation function. Moreover, we optimized the neural network by means of cross entropy loss function using the \textit{RMSProp} optimization algorithm. Figure \subref{fig:classification} shows the performance of the different Neural Networks. Observe that CNN has the highest Macro F1 score in both data sets, achieving 99\% in the training set and 91\% in the test set, followed by LSTM with 93\% in the training set and 89\% in the test set. Finally, the FastText neural network achieved a 86\% Macro F1 score in the training set and 88\% in the test set. 

Finally, in order to verify if there is any significant bias in our results, we evaluated the accuracy of the classification for each deputy and each month separately. Figure \subref{fig:accuracy} shows the Macro F1 score per deputy and per month. Observe that the median Macro F1 score for deputies is $0.95$, with a minimum of $0.84$. 
Similarly, the results per month were also good in general, with a median Macro F1 score of $0.97$ and a minimum of $0.83$. For both cases, outliers obtained better results than a random classifier. 

\section{Qualitative Evaluation}
\label{sec:results}

\begin{figure}[h]
\centering
\begin{subfigure}[b]{0.23\textwidth}
\includegraphics[width=\linewidth]{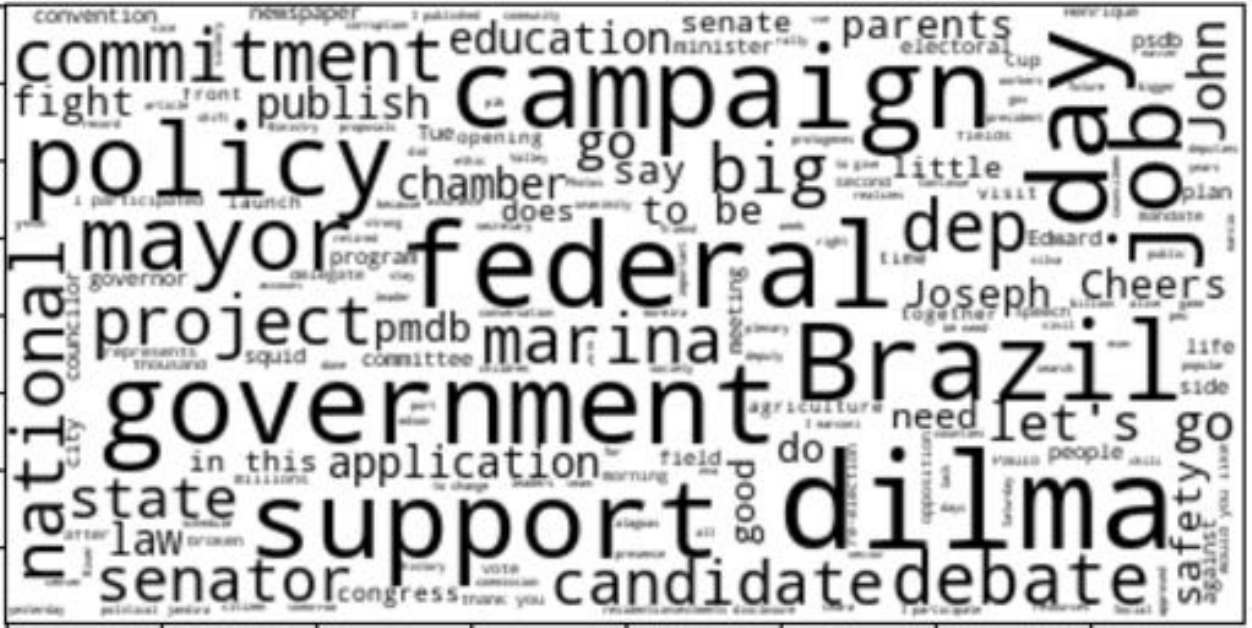}
\label{fig:cloud_e_p}
\end{subfigure}
\begin{subfigure}[b]{0.23\textwidth}
\includegraphics[width=\linewidth]{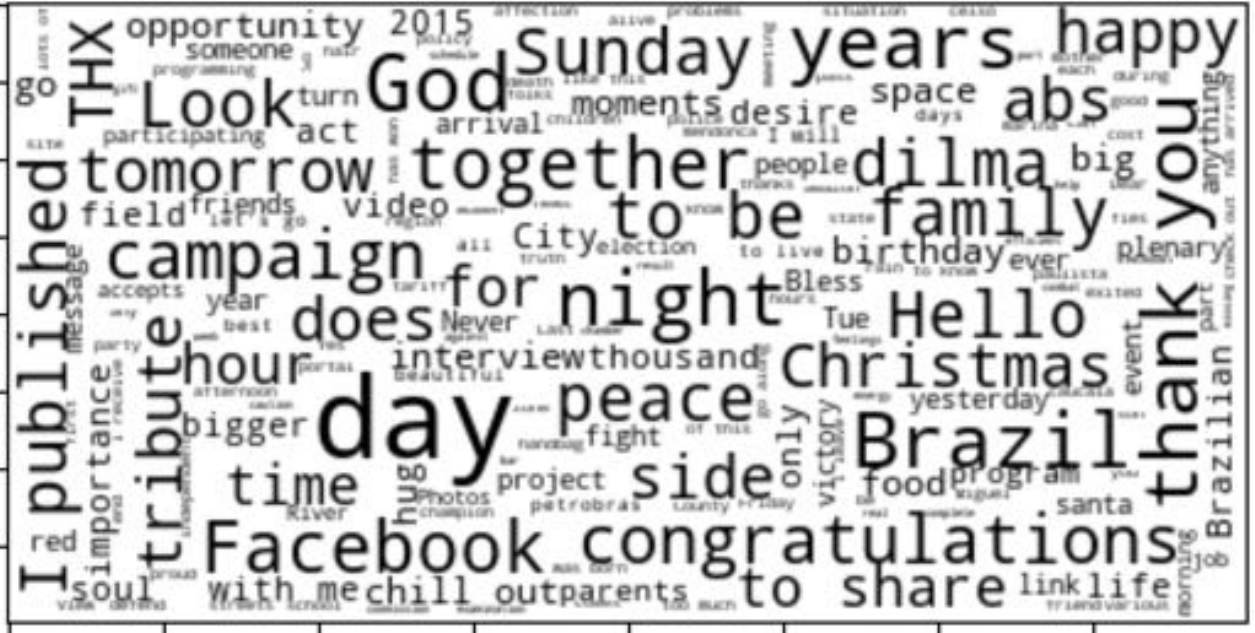}
\label{fig:cloud_e_np}
\end{subfigure}
\vspace{-0.3cm}
\caption{Word clouds of political (left) and non-political (right) tweets during elections.}
\label{fig:cloud}
\end{figure}

Besides the Macro F1 score, we also evaluated our proposal from qualitative perspective\footnote{All results were translated from Portuguese.}. 
First, we used word clouds, a visual representation that depicts the most frequent words in a text. Our objective is to visualize the most common political and non-political words that occurs in congressmen tweets in election period. 
Figure \ref{fig:cloud} exhibits the word clouds of political and non-political tweets during this period. Figure~\ref{fig:cloud}a shows the main words related to politics in election period. The most preeminent words are "campaign","federal","support","government" and "dilma", which clearly indicate campaign tweets. In contrast, Figure~\ref{fig:cloud}b shows the most predominant words of non-political tweets for the same period. Observe that these words could be related to personalism and/or clientelism/patronage, e.g."God", "I published", "Facebook", "together", "congratulations", "thank you", etc.

Second, we used a short text topic modeling approach, namely BTM (Biterm Topic Modeling) \cite{Yan2013}, to identify the main political and non-political latent topics in parliamentarian tweets.
Using our classification method, we first separate all tweets into two disjoint corpora: political and non-political tweets. After that, we trained the BTM algorithm with these two corpora, using the following parameters: number of topics $K = 10$, $\alpha = 50/K$, $\beta = 0.005$ and number of iterations = 1000.

Results are shown in Table \ref{topics}, which lists the five most important words of each topic for both corpora. The use of BTM jointly with our method highlighted the use of personalism by Brazilian parliamentarians. In non-political topics, note that there are many personalistic topics (2, 3, 5, 6, 8, 9, 10), which are represented by words such as ``congratulations'', ``friend'', ``hugs'', ``moments'' and ``together''. Concerning the political corpus, observe that there are indications of patronage or a more pragmatic behavior of politicians, such as (2, 5) and others related to the schedule of the deputies (6, 7, 10). Since the classification is not perfect, we found a personalistic topic (8) in the middle of the political tweet corpus. Nevertheless, the other topics reflect important facts and issues related to the Brazilian political scenario, such as the political crisis (1, 3). 

What is striking about our analysis is the amount of non-political discourse presented in the tweets of Brazilian deputies. Observe in Table~\ref{tab:data} that at least half of tweets are non-political for all classes of deputies. Concerning the specific topics, the two most popular topics, comprising 25\% of all tweets, are non-political: topics (1, 4), of a broad and populist nature. Note also that political tweets are scattered over many topics and the most important is topic (6), which is related to pragmatic deputy behavior and comprises 8\% of all tweets.

\begin{table}[h!]
\centering

\caption{The 5 most important words from 10 topics of the political and non-political tweet corpora.}
\label{topics}
\resizebox{\linewidth}{!}{
\begin{tabular}{llr|lr}
\hline
\# & \textbf{Non-Political Topics} & \textbf{\% tweets} &\textbf{Political Topics} & \textbf{\% tweets} \\ \hline
1	&	brazil can be country go	&	12	&	government dilma against politics brazil	&	5	\\
2	&	today city big day friends	&	5	&	mayor deputy today president meeting	&	3	\\
3	&	day all god friends week	&	3	&	dilma lula temer against president	&	4	\\
4	&	thousand brazil millions go project	&	6	&	meeting about today national chamber	&	8	\\
5	&	for friend big brazil today	&	5	&	health federal government resources state	&	4	\\
6	&	day today brazil years congratulations	&	13	&	chamber commission deputies pec project	&	3	\\
7	&	program live day radio interview	&	2	&	about law against education project	&	5	\\
8	&	together hugs see moments relaxation	&	1	&	day today every year god	&	3	\\
9	&	facebook posted photos album posted	&	5	&	government dilma brazil year millions	&	6	\\
10	&	food soul grace lauds los	&	1	&	deputy about federal today chamber	&	6	\\ \hline
\end{tabular}
}
\end{table}

\section{Conclusions}
\label{sec:conclusions}
In this work we proposed a supervised method to identify short messages with political and non-political content posted by parliamentarians. Quantitative and qualitative results showed that our proposal was able to accurately separate political from non-political tweets.
\footnotesize

\begin{thebibliography}{}

\bibitem[\protect\citeauthoryear{Amaral and Pinho}{2016}]{amaral2016Tuitando}
Amaral, M.~S., and Pinho, J. A. G.~d.
\newblock 2016.
\newblock Tuitando por votos: Congressistas brasileiros e o uso do twitter nas
  eleições de 2014.
\newblock In {\em Proceedings of the XL Encontro da Anpad},  1--19.
\newblock Anpad.

\bibitem[\protect\citeauthoryear{Bucci}{2016}]{BUCCI2016}
Bucci, E.
\newblock 2016.
\newblock {\em {A forma bruta dos protestos: das manifesta{\c{c}}{\~{o}}es de
  junho de 2013 {\`{a}} queda de Dilma Rousseff em 2016}}.
\newblock S{\~{a}}o Paulo: Cia das Letras.

\bibitem[\protect\citeauthoryear{Conover, Ratkiewicz, and
  Francisco}{2011}]{Conover2011}
Conover, M.; Ratkiewicz, J.; and Francisco, M.
\newblock 2011.
\newblock {Political polarization on twitter.}
\newblock {\em Icwsm} 133(26):89--96.

\bibitem[\protect\citeauthoryear{DiGrazia \bgroup et al\mbox.\egroup
  }{2013}]{DiGrazia2013}
DiGrazia, J.; McKelvey, K.; Bollen, J.; and Rojas, F.
\newblock 2013.
\newblock {More tweets, more votes: Social media as a quantitative indicator of
  political behavior}.
\newblock {\em PLoS ONE} 8(11):1--5.

\bibitem[\protect\citeauthoryear{Downie}{2014}]{Downie}
Downie, A.
\newblock 2014.
\newblock {Brazil crowds up but stadium usage still problematic}.

\bibitem[\protect\citeauthoryear{Duchi, Hazan, and Singer}{2011}]{Duchi2011}
Duchi, J.; Hazan, E.; and Singer, Y.
\newblock 2011.
\newblock {Adaptive Subgradient Methods for Online Learning and Stochastic
  Optimization}.
\newblock {\em Journal of Machine Learning Research} 12:2121--2159.

\bibitem[\protect\citeauthoryear{Gao, Kuppersmith, and Huang}{2017}]{Gao2017}
Gao, L.; Kuppersmith, A.; and Huang, R.
\newblock 2017.
\newblock {Recognizing Explicit and Implicit Hate Speech Using a Weakly
  Supervised Two-path Bootstrapping Approach}.

\bibitem[\protect\citeauthoryear{Hartmann \bgroup et al\mbox.\egroup
  }{2017}]{Hartmann2017}
Hartmann, N.; Fonseca, E.; Shulby, C.; Treviso, M.; Rodrigues, J.; and Aluisio,
  S.
\newblock 2017.
\newblock {Portuguese Word Embeddings: Evaluating on Word Analogies and Natural
  Language Tasks}.
\newblock (Section 3).

\bibitem[\protect\citeauthoryear{Hemphill, Otterbacher, and
  Shapiro}{2013}]{hemphill2013s}
Hemphill, L.; Otterbacher, J.; and Shapiro, M.
\newblock 2013.
\newblock What's congress doing on twitter?
\newblock In {\em Proceedings of the 2013 conference on Computer supported
  cooperative work},  877--886.
\newblock ACM.

\bibitem[\protect\citeauthoryear{Hochreiter and Schmidhuber}{1997}]{Hochreiter}
Hochreiter, S., and Schmidhuber, J.
\newblock 1997.
\newblock Long short-term memory.
\newblock {\em Neural Comput.} 9(8):1735--1780.

\bibitem[\protect\citeauthoryear{Joulin \bgroup et al\mbox.\egroup
  }{2016}]{Joulin2016}
Joulin, A.; Grave, E.; Bojanowski, P.; and Mikolov, T.
\newblock 2016.
\newblock {Bag of Tricks for Efficient Text Classification}.

\bibitem[\protect\citeauthoryear{Kim}{2014}]{Kim2014}
Kim, Y.
\newblock 2014.
\newblock {Convolutional Neural Networks for Sentence Classification}.
\newblock  1746--1751.

\bibitem[\protect\citeauthoryear{Le and Mikolov}{2014}]{le2014distributed}
Le, Q., and Mikolov, T.
\newblock 2014.
\newblock Distributed representations of sentences and documents.
\newblock In {\em Proceedings of the 31st International Conference on Machine
  Learning (ICML-14)},  1188--1196.

\bibitem[\protect\citeauthoryear{Lietz \bgroup et al\mbox.\egroup
  }{2014}]{Lietz2014}
Lietz, H.; Wagner, C.; Bleier, A.; and Strohmaier, M.
\newblock 2014.
\newblock {When Politicians Talk: Assessing Online Conversational Practices of
  Political Parties on Twitter}.
\newblock {\em Eighth International AAAI Conference on Weblogs and Social
  Media}  285--294.

\bibitem[\protect\citeauthoryear{Mainwaring}{2001}]{mainwaring2001sistemas}
Mainwaring, S.
\newblock 2001.
\newblock {\em Sistemas partid{\'a}rios em novas democracias: o caso do
  Brasil}.
\newblock Mercado Aberto.

\bibitem[\protect\citeauthoryear{Mikolov \bgroup et al\mbox.\egroup
  }{2013}]{Mikolov2013}
Mikolov, T.; Chen, K.; Corrado, G.; and Dean, J.
\newblock 2013.
\newblock {Efficient Estimation of Word Representations in Vector Space}.
\newblock {\em Arxiv}  1--12.

\bibitem[\protect\citeauthoryear{Mois{\'e}s}{2011}]{moises2011desempenho}
Mois{\'e}s, J.~{\'A}.
\newblock 2011.
\newblock O desempenho do congresso nacional no presidencialismo de
  coaliz{\~a}o (1995-2006).
\newblock In {\em O papel do Congresso Nacional no presidencialismo de
  coaliz{\~a}o}.

\bibitem[\protect\citeauthoryear{Nogueira}{2013}]{nogueira2013ruas}
Nogueira, M.~A.
\newblock 2013.
\newblock {\em As ruas ea democracia: ensaios sobre o Brasil
  contempor{\^a}neo}.
\newblock Funda{\c{c}}{\~a}o Astrojildo Pereira.

\bibitem[\protect\citeauthoryear{Panagiotopoulos and
  Sams}{2012}]{panagiotopoulos2012overview}
Panagiotopoulos, P., and Sams, S.
\newblock 2012.
\newblock An overview study of twitter in the uk local government.

\bibitem[\protect\citeauthoryear{Paul \bgroup et al\mbox.\egroup }{2017}]{Paul}
Paul, D.; Li, F.; Teja, M.~K.; Yu, X.; and Frost, R.
\newblock 2017.
\newblock {Compass: Spatio Temporal Sentiment Analysis of US Election}.
\newblock In {\em Proceedings of the 23rd ACM SIGKDD International Conference
  on Knowledge Discovery and Data Mining - KDD '17},  1585--1594.
\newblock New York, New York, USA: ACM Press.

\bibitem[\protect\citeauthoryear{Pennington, Socher, and
  Manning}{2014}]{pennington2014glove}
Pennington, J.; Socher, R.; and Manning, C.
\newblock 2014.
\newblock Glove: Global vectors for word representation.
\newblock In {\em Proceedings of the 2014 conference on empirical methods in
  natural language processing (EMNLP)},  1532--1543.

\bibitem[\protect\citeauthoryear{Rori and Richards}{2017}]{Rori2017}
Rori, L., and Richards, B.
\newblock 2017.
\newblock {\em {Understanding Online Political Networks: The case of the far
  right and far left in Greece Pantelis}}, volume 10540 of {\em Lecture Notes
  in Computer Science}.
\newblock Cham: Springer International Publishing.

\bibitem[\protect\citeauthoryear{Saad-Filho}{2013}]{saad2013mass}
Saad-Filho, A.
\newblock 2013.
\newblock Mass protests under ‘left neoliberalism’: Brazil, june-july 2013.
\newblock {\em Critical Sociology} 39(5):657--669.

\bibitem[\protect\citeauthoryear{Samuels and Zucco}{2014}]{samuels2014lulismo}
Samuels, D.~J., and Zucco, C.
\newblock 2014.
\newblock Lulismo, petismo, and the future of brazilian politics.

\bibitem[\protect\citeauthoryear{Samuels}{1999}]{samuels1999incentives}
Samuels, D.~J.
\newblock 1999.
\newblock Incentives to cultivate a party vote in candidate-centric electoral
  systems: Evidence from brazil.
\newblock {\em Comparative Political Studies} 32(4):487--518.

\bibitem[\protect\citeauthoryear{Tumasjan \bgroup et al\mbox.\egroup
  }{2010}]{tumasjan2010predicting}
Tumasjan, A.; Sprenger, T.~O.; Sandner, P.~G.; and Welpe, I.~M.
\newblock 2010.
\newblock {Predicting Elections with Twitter: What 140 Characters Reveal about
  Political Sentiment.}
\newblock {\em ICWSM} 10:178--185.

\bibitem[\protect\citeauthoryear{Watts}{2016}]{Watts2016}
Watts, J.
\newblock 2016.
\newblock {Dilma Rousseff impeachment: what you need to know – the Guardian
  briefing}.

\bibitem[\protect\citeauthoryear{Yan \bgroup et al\mbox.\egroup
  }{2013}]{Yan2013}
Yan, X.; Guo, J.; Lan, Y.; and Cheng, X.
\newblock 2013.
\newblock {A Biterm Topic Model for Short Texts}.
\newblock In {\em Proceedings of the 22Nd International Conference on World
  Wide Web},  1445--1455.

\end{thebibliography}

\bibliographystyle{aaai}
\end{document}